\documentclass[aps,prl,twocolumn,groupedaddress,showpacs]{revtex4-1}
\usepackage{graphicx}
\usepackage{amssymb}
\usepackage{amsmath}

\begin{document}
\title{Hydrodynamics with spin in bacterial suspensions}
\author{M.Belovs}
\affiliation{University of Latvia, Ze\c{l}\c{l}u-23, R\={\i}ga, LV-1002, Latvia}
\author{A.C\={e}bers}
\email[]{aceb@tesla.sal.lv}
\affiliation{University of Latvia, Ze\c{l}\c{l}u-23, R\={\i}ga, LV-1002, Latvia}

\begin{abstract}
We describe a new kind of self-propelling motion of bacteria based on the cooperative action of rotating flagella on the surface of bacteria. Describing the ensemble of rotating flagella in the framework of the hydrodynamics with spin the reciprocal theorem of Stokesian hydrodynamics is generalized accordingly. The velocity of the self-propulsion is expressed in terms of the characteristics of the vector field of flagella orientation and  it is shown that unusually high velocities of \textit{Thiovulum majus} bacteria may be explained by the cooperative action of the rotating flagella. The expressions obtained enable us to estimate the torque created by the rotary motors of the bacterium and show quantitative agreement with the existing experimental data.
\end{abstract}


\maketitle

Systems where an important role is played by the rotation of constituent particles are well known. Particular examples include liquid crystals \cite{1}, magnetic liquids \cite{2} and others. The concept of a viscous stress induced by particle rotations allows one to describe such phenomena as the increase of the effective viscosity of a magnetic fluid in an applied magnetic field \cite{3}, the rotation of magnetic liquid droplets in a rotating field \cite{4}, the macroscopic flows caused by internal  rotations on the free boundaries of a liquid sample \cite{5} and negative effective viscosity of so-called Quincke suspensions \cite{6}.

It was recently  found that macroscopic flow may arise in bacterial suspensions due to the rotation of bacteria \cite{7}. The bacteria \textit{Thiovulum majus} were shown to form rotating aggregates near solid walls which may be interpreted as rotating crystals \cite{7}. It is important to note that \textit{Thiovulum majus} bacteria are among the fastest microorganisms and their velocity of self-propelling motion may reach $600~\mu m/s$ - an order of magnitude faster than, for example, the well-known bacteria \textit{E.coli} \cite{8}. As noted in \cite{9} the physical mechanism behind the high self-propulsion velocity of \textit{Thiovulum majus} bacteria is not known at present.

Here we show how the concept of stress induced by torques acting on flagella of a \textit{Thiovulum majus} bacterium enables us to explain this phenomenon. We start by generalizing the reciprocal theorem of the Stokesian hydrodynamics \cite{10} to the case of a fluid with antisymmetric viscous stresses.

\begin{figure}
\includegraphics[width=1.0\columnwidth]{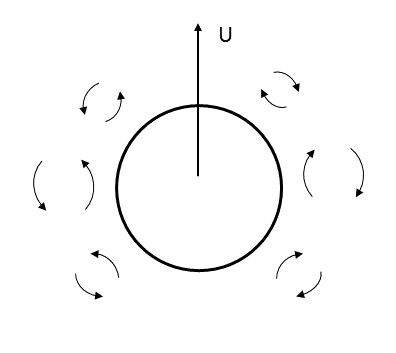}
\caption{Sketch of self-propulsion caused by rotations of flagella.}
\label{fig1}
\end{figure}

The equation of motion for a liquid with antisymmetric viscous stresses reads
\begin{equation}
-\nabla p+\nabla\cdot\vec{\sigma}^{s}+\nabla\cdot\vec{\sigma}^{a}=\nabla\cdot\vec{\sigma}=0;~div(\vec{u})=0~,
\label{Eq:1}
\end{equation}
where $\sigma^{s}_{ik}=\eta\Bigl(\frac{\partial u_{i}}{\partial x_{k}}+\frac{\partial u_{k}}{\partial x_{i}}\Bigr)$, $\sigma^{a}_{ik}=\frac{1}{2}e_{ikl}\tau_{l}$, $\vec{\tau}$ is the volume density of torques acting on the liquid. Since \textit{Thiovulum majus} bacteria have a large quantity of short rotating flagella (on the order of several hundred according to \cite{7,9}) we take into account torques in a liquid layer of thickness $L$ that surrounds the bacterium. Besides the flow and stress of the liquid with spin given  by $(\vec{u},\vec{\sigma})$ according to Eq.(\ref{Eq:1}) we also consider the solution of the Stokes equation
for the flow around a sphere moving with velocity $\hat{\vec{U}}$ or rotating with an angular velocity $\hat{\vec{\Omega}}$
\begin{equation}
-\nabla \hat{p}+\nabla\cdot\hat{\vec{\sigma}}^{s}=\nabla\cdot\hat{\vec{\sigma}}=0;~div(\hat{\vec{u}})=0~.
\label{Eq:2}
\end{equation}
The flow velocity and stress obtained by solving  Eq.(\ref{Eq:2}) are denoted $(\hat{\vec{u}},\hat{\vec{\sigma}})$.
Multiplying the equation of motion $\frac{\partial \sigma_{ik}}{\partial x_{k}}=0$ by $\hat{u}_{i}$, integrating by parts and  using Eqs.(\ref{Eq:1},\ref{Eq:2})  and the condition of incompressibility we obtain the following generalization of the reciprocal theorem in the case of the liquid with spin ($\vec{U}$ is the velocity of a sphere in the liquid described by Eq.(\ref{Eq:1}), $\vec{n}$ is the normal to the surface of sphere, surface integrals are taken along the surface of sphere, the volume integral is taken along the region occupied by the surrounding liquid):
\begin{equation}
\int \hat{U}_{i}\sigma_{ik}n_{k}dS=\int U_{i}\hat{\sigma}_{ik}n_{k}dS+\int\vec{\tau}\hat{\vec{\Omega}}_{0}dV~,
\label{Eq:3}
\end{equation}
where $\hat{\vec{\Omega}}_{0}=\nabla\times\hat{\vec{u}}/2$.
Let us consider the case when $\hat{\vec{u}},\hat{\vec{\sigma}}$ correspond to the velocity and stress fields in the case of sphere moving with a constant velocity $\hat{\vec{U}}$. According to the solution of the Stokes problem $\hat{\sigma}_{ik}n_{k}=-\frac{3\eta}{2a}\hat{U}_{i}$ and $\hat{\vec{\Omega}}_{0}=\frac{3a}{4}\frac{\hat{\vec{U}}\times\vec{r}}{r^{3}}$. Since the self-propelling motion with velocity $\vec{U}$ occurs at zero force $\int \sigma_{ik}n_{k}dS=0$ then the generalized reciprocal theorem (\ref{Eq:3}) yields
\begin{equation}
\vec{U}=\frac{1}{8\pi\eta}\int \frac{\vec{r}\times\vec{\tau}}{r^{3}}dV~.
\label{Eq:4}
\end{equation}
It is therefore the distribution of torques in the liquid layer surrounding the sphere that causes the self-propelling motion. Since the flagella of \textit{Thiovulum majus} are short we can approximate the relation (\ref{Eq:4}) for the velocity of self-propulsion in the direction of the unit vector $\vec{e}_{u}$ as follows ($\vec{e}_{r}=\vec{r}/r;~\vec{e}_{\tau}=\vec{\tau}_{f}/\tau_{f}$) ($d\omega$ is the area of the surface element of the sphere with the radius equal to one)
\begin{equation}
\vec{U}\cdot\vec{e}_{u}=\frac{4\pi nL\tau_{f}}{8\pi\eta}\frac{1}{4\pi}\int \vec{e}_{u}\times \vec{e}_{r}\cdot\vec{e}_{\tau}d\omega~,
\label{Eq:5}
\end{equation}
where $\tau_{f}$ is the torque created by a single flagellum and $nL$ is the number of flagella per unit surface area of the bacterium. Since $nL=N/(4\pi a^{2})$ (where $N$ denotes the number of flagella per bacterium) the relation (\ref{Eq:5}) may be rewritten as follows
\begin{equation}
\vec{U}\cdot\vec{e}_{u}=\frac{N\tau_{f}}{8\pi\eta a^{2}}k_{u}~,
\label{Eq:6}
\end{equation}
where we have introduced the notation
\begin{equation}
k_{u}= \frac{1}{4\pi}\int \vec{e}_{u}\times \vec{e}_{r}\cdot\vec{e}_{\tau}~.
\label{Eq:7}
\end{equation}
We see that the self-propulsion velocity depends on the vector field $\vec{e}_{\tau}$, which may be 
expressed in terms of vector spherical harmonics \cite{11} $\vec{Y}_{lm}=\vec{e}_{r}Y_{lm};\vec{\Psi}_{lm}=\nabla Y_{lm};\vec{\Phi}_{lm}=(\vec{e}_{r}\times\nabla)Y_{lm}$ (where $Y_{lm}$ denotes the conventional spherical harmonics and the operator $\nabla=(0,\frac{\partial}{\partial \vartheta},\frac{1}{\sin{(\vartheta)}}\frac{\partial}{\partial\varphi}))$ as follows
\begin{equation}
\vec{e}_{\tau}=\sum_{l,m}(a_{lm}\vec{Y}_{lm}+b_{lm}\vec{\Phi}_{lm}+c_{lm}\vec{\Psi}_{lm})~.
\label{Eq:7a}
\end{equation}
Expressing $\vec{e}_{u}=\cos{(\vartheta)}\vec{e}_{r}-\sin{(\vartheta)}\vec{e}_{\vartheta}$, using $\vec{\Psi}_{10}=-\sqrt{\frac{3}{4\pi}}\sin{(\vartheta)}\vec{e}_{\vartheta}$ and the orthogonality of vector spherical harmonics
\begin{equation}
\int\vec{\Psi}_{lm}\vec{\Psi}^{*}_{l'm'}d\omega=l(l+1)\delta_{ll'}\delta_{mm'}
\label{Eq:7b}
\end{equation}
we may express $k_{u}$ as $k_{u}=-b_{10}/\sqrt{3\pi}$ which gives the azimuthal component of the vector field $\vec{e}_{\tau}=\frac{3}{2}k_{u}\sin{(\vartheta)}\vec{e}_{\varphi}$. Thus for self-propulsion it is necessary that the vector field $\vec{e}_{\tau}$ has azimuthal component which creates thrust as shown by the sketch in Fig.~\ref{fig1}.
In some sense a similar thing happens for \textit{Volvox} algae cells, which flagella beats in a plane that is tilted approximately $15^{o}$ from the anterior-posterior axis causing their rotation \cite{12}. Angular dependence of the azimuthal component of $\vec{e}_{\tau}\sim\sin{(\vartheta)}$ which gives the contribution to the thrust is similar to the slip velocity dependence of the celebrated squirmer model \cite{13}.
The corresponding calculation in the frame of the boundary layer approach confirms the result given by Eq.(\ref{Eq:6}).

The trajectory of a \textit{Thiovulum majus} bacterium  is a helix \cite{14} which means that in addition to the self-propulsion the bacterium also rotates around the axis which makes an angle with the direction of the velocity. The reciprocal theorem (\ref{Eq:3}) allows us to obtain the relation for the angular velocity of the bacterium rotation. Taking $\hat{\vec{U}}=\hat{\vec{\Omega}}\times\vec{r}$ on the surface of a sphere rotating with the angular velocity $\hat{\vec{\Omega}}$, when $\hat{\vec{\sigma}}\cdot\vec{n}=-3\eta\hat{\vec{\Omega}}\times\vec{n}$ and $\vec{\Omega}_{0}=\frac{a^{3}}{2}\Bigl(\frac{3\vec{r}(\hat{\vec{\Omega}}\cdot\vec{r})}{r^{5}}-\frac{\hat{\vec{\Omega}}}{r^{3}}\Bigr)$,  and applying the condition of zero torque relation (\ref{Eq:3}) gives
\begin{equation}
\Omega_{i}=\frac{1}{16\pi\eta}\int \Bigl(\frac{3r_{i}(\vec{\tau}\cdot\vec{r})}{r^{5}}-\frac{\tau_{i}}{r^{3}}\Bigr)~.
\label{Eq:8}
\end{equation}
Estimating the right side of Eq.(\ref{Eq:8}) as previously for the angular velocity of bacterium rotation we have ($\vec{e}_{\Omega}$ is the unit vector along the angular velocity)
\begin{equation}
\vec{e}_{\Omega}\cdot\vec{\Omega}=\frac{N\tau_{f}}{16\pi\eta a^{3}}k_{\Omega}~,
\label{Eq:9}
\end{equation}
where we have introduced the notation 
\begin{equation}
k_{\Omega}=\frac{1}{4\pi}\int\Bigl(3(\vec{e}_{r}\cdot\vec{e}_{\tau})(\vec{e}_{\Omega}\cdot\vec{e}_{r})-(\vec{e}_{\Omega}\cdot\vec{e}_{\tau}\Bigr)d\omega~.
\label{Eq:10}
\end{equation}
As a result we can obtain a simple relation between the angular velocity of the bacterium and its velocity of self-propulsion
\begin{equation}
U=2a\Omega\frac{k_{u}}{k_{\Omega}}~.
\label{Eq:11}
\end{equation}
Taking $a=8.5 ~\mu m$, $\Omega=50~s^{-1}$ \cite{7} for the angular velocity we can obtain an estimate for the velocity of self-propulsion $U\simeq 850~\mu m/s$ - a value close to the experimental findings. Thus, the rapid self-propulsion velocity of \textit{Thiovulum majus} may be explained by the cooperative action of the ensemble of rotating flagella distributed along the surface of the bacterium. The coefficients $k_{u},k_{\Omega}$ depend on the distribution of the flagella on the surface of bacterium about which, as far as we know, there is little information at the present time.

Eq.(\ref{Eq:6}) allows us to estimate the torque created by the rotary motors of bacterium $\tau_{f}$. Taking $U=600~\mu m/s;~a=8.5~\mu m;~\eta=1~cP$ and $N=200$ we obtain $\tau_{f}\simeq 5~pN\cdot\mu m$ - a value quite close to results obtained experimentally by different means and referenced in \cite{12}. This gives additional evidence that the unusually high velocity of self-propulsion of \textit{Thiovulum majus} bacteria is caused by the cooperative motion in the ensembles of rotating flagella distributed along the surfaces of bacteria.

In \cite{7} the formation of rotating crystals of \textit{Thiovulum majus} bacteria is described and data for the angular velocity of the crystal in dependence on its size are given. Since the mechanism of this phenomenon is similar to the considered in the first part we will further describe it within the framework of the hydrodynamics with spin. Bacteria stuck to a solid wall due to the flow caused by the presence of solid boundary form a dense aggregate of rotating bacteria \cite{7}. The volumetric torque density in the liquid may be estimated according to the relation $n\alpha\Omega_{0}$ ($\alpha=8\pi\eta a^{3}$), where $n$ is number of bacteria per unit volume $a$ is their radius and $\Omega_{0}$ ($5<\Omega_{0}<10^{2}~s^{-1}$) is the angular velocity of their rotation. The Stokes equation for the cylindrical layer of bacteria with radius $R$ and thickness $h$ surrounded by liquid with the same viscosity under the action of torques caused by bacteria reads
\begin{equation}
-\nabla p+\eta\Delta\vec{v}+\frac{1}{2}\nabla\times\vec{\tau}=0;~div\vec{v}=0~,
\label{Eq:12}
\end{equation}
where $\vec{\tau}=n\alpha\Omega_{0}\vec{e}_{z}\theta(R-r)\theta(h-z)$ which should be solved in the half-space $z>0$ at the no-slip boundary condition $\vec{v}\mid_{z=0}=0$. The torques induce the azimuthal force on the boundary of the aggregate
$\vec{f}=\vec{e}_{\varphi}n\alpha\Omega_{0}\delta(R-r)\theta(h-z)/2$, which causes a rotational flow in the direction of the bacterial rotation. In that sense the behavior of the aggregate is similar to the behavior of magnetic fluid droplet in a rotating magnetic field \cite{4} or the less trivial case of an ensemble of chiral particles \cite{13}. The solution of the problem is found by the 2D Fourier transform
\begin{equation}
\vec{v}(x,y,z)=\frac{1}{2\pi}\int\vec{v}_{\vec{k}}(z)\exp{(ik_{x}x+ik_{y}y)}dk_{x}dk_{y}
\end{equation}
and
\begin{equation}
p(x,y,z)=\frac{1}{2\pi}\int p_{\vec{k}}(z)\exp{(ik_{x}x+ik_{y}y)}dk_{x}dk_{y}~,
\end{equation}
which after lengthy but straightforward calculations yields the following expression  for the azimuthal velocity of the flow 
\begin{widetext}
\begin{equation}
v_{\varphi}=\frac{n\alpha\Omega_{0}R}{2\eta}\int^{\infty}_{0}\frac{J_{1}(kR)J_{1}(kr)}{k}(1-\exp{(-kz)}-\exp{(-kh)}\sinh{(kz)})dk~(z<h)~.
\label{Eq:15}
\end{equation}
For the period of the aggregate rotation $T=2\pi R/v_{\varphi}(R,h)$ this yields
\begin{equation}
T=\frac{4}{n(2a)^{3}\Omega_{0}}\Bigl(\int^{\infty}_{0}\frac{J_{1}^{2}(u)}{u}(1-\exp{(-uh/R)}-\exp{(-uh/R)}\sinh{(uh/R)})du\Bigr)^{-1}
\label{Eq:16}
\end{equation}
\end{widetext}
The multiplier  $4/(n(2a)^{3}\Omega_{0})$ is estimated as $2/50~s$ assuming a dense packing of bacteria in the layer. 
\begin{figure}
\includegraphics[width=1.0\columnwidth]{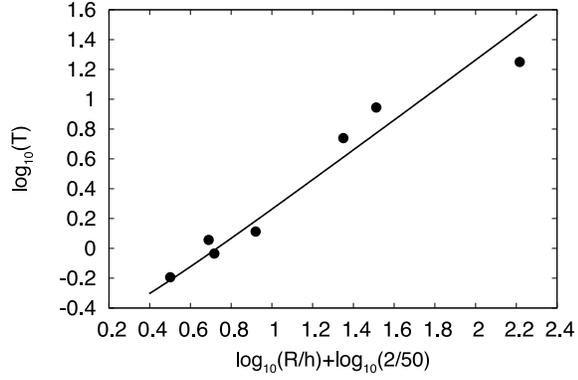}
\caption{Period of the crystal rotation T (in seconds) as a function of its size. $4\pi\eta/(n\alpha\Omega_{0})=2/50~s$. Solid line calculated according to relation (\ref{Eq:16}). Circles experimental data from \cite{7}.}
\label{fig2}
\end{figure}
Fig.~\ref{fig2} shows the period $T$ as a function of the ratio $R/h$ in logarithmic coordinates to match the experimental data \cite{7}. We see that the model of the liquid with spin predicts linear growth of the period of aggregate rotation with its size, which is in full agreement with the  experimental data.

In summary, we have described evidence that the approach of the hydrodynamics with spin is fruitful for bacterial hydrodynamics. It allows us to explain the unusually high velocities of self-propulsion of \textit{Thiovulum majus} bacteria and describe the flow arising in aggregates of rotating bacteria. 
\section{Ackowledgements} Authors are thankful to R.Livanovics for fruitful discussions. A.C. acknowledges support from National Research Programme No.2014.10-4/VPP-3/21.

\end{document}